\documentclass[letterpaper]{article} 
\usepackage{aaai24}  
\usepackage{times}  
\usepackage{helvet}  
\usepackage{courier}  
\usepackage[hyphens]{url}  
\usepackage{graphicx} 
\usepackage{natbib}  
\usepackage{caption} 
\usepackage{algorithm}
\usepackage{algorithmicx}
\usepackage{algpseudocode}
\usepackage{adjustbox}
\usepackage{booktabs} 
\usepackage{multirow}
\usepackage{enumerate}
\usepackage{amssymb}
\usepackage{amsmath}
\usepackage{dsfont}
\newcommand{\tabincell}[2]{\begin{tabular}{@{}c#1@{}}#2\end{tabular}} 
\usepackage{newfloat}
\usepackage{listings}
\DeclareCaptionStyle{ruled}{labelfont=normalfont,labelsep=colon,strut=off} 
\floatstyle{ruled}
\newfloat{listing}{tb}{lst}{}
\floatname{listing}{Listing}
\title{Progressive Poisoned Data Isolation for Training-time Backdoor Defense}
\author{
 Yiming Chen, 
 Haiwei Wu, and 
 Jiantao Zhou\textsuperscript{\footnotemark[2]}
}
\affiliations{
State Key Laboratory of Internet of Things for Smart City\\
 Department of Computer and Information Science, University of Macau\\
 {\tt\small \{yc17486, yc07912, jtzhou\}@umac.mo}
}

\begin{document}

\maketitle
\renewcommand{\thefootnote}{\fnsymbol{footnote}} 
\footnotetext[2]{Corresponding author.} 

\begin{abstract}
\indent Deep Neural Networks (DNN) are susceptible to backdoor attacks where malicious attackers manipulate the model's predictions via data poisoning. It is hence imperative to develop a strategy for training a clean model using a potentially poisoned dataset. Previous training-time defense mechanisms typically employ an one-time isolation process, often leading to suboptimal isolation outcomes. In this study, we present a novel and efficacious defense method, termed Progressive Isolation of Poisoned Data (PIPD), that progressively isolates poisoned data to enhance the isolation accuracy and mitigate the risk of benign samples being misclassified as poisoned ones. Once the poisoned portion of the dataset has been identified, we introduce a selective training process to train a clean model. Through the implementation of these techniques, we ensure that the trained model manifests a significantly diminished attack success rate against the poisoned data. Extensive experiments on multiple benchmark datasets and DNN models, assessed against nine state-of-the-art backdoor attacks, demonstrate the superior performance of our PIPD method for backdoor defense. For instance, our PIPD achieves an average True Positive Rate (TPR) of 99.95\% and an average False Positive Rate (FPR) of 0.06\% for diverse attacks over CIFAR-10 dataset, markedly surpassing the performance of state-of-the-art methods. The code is available at https://github.com/RorschachChen/PIPD.git.

\end{abstract}

\section{Introduction}

The utilization of diverse datasets in training DNNs enhances their adaptability and performance across various tasks and domains. However, the demand for multiple sources of data also introduces a vulnerability to backdoor attacks \cite{BadNets}. Malicious attackers can exploit this situation by injecting hidden backdoors into the training data, thereby manipulating the predictions of the model. The potential harm of backdoor attacks lies in their ability to trigger malicious behaviors in the deployed model once activated. Such attacks can lead to the disruption of system operations, and even system crashes. 

Backdoor attack is a type of adversary \cite{NEURIPS2020_8b406655} that implants malicious backdoor behaviors into the victim models during training. This goal can be accomplished by introducing contamination into the data used for training. The existing backdoor attacks can be roughly divided into two categories \cite{beagle}, namely \emph{patching} and \emph{transforming}. Specifically, in patching attacks, a trigger is injected to a benign sample by merging their pixel values with a mask. In contrast, transforming attacks inject the trigger into the input using a transformation function in the form of an algorithm or a network. 
\begin{figure}[t]
\centering
\includegraphics[width=\columnwidth]{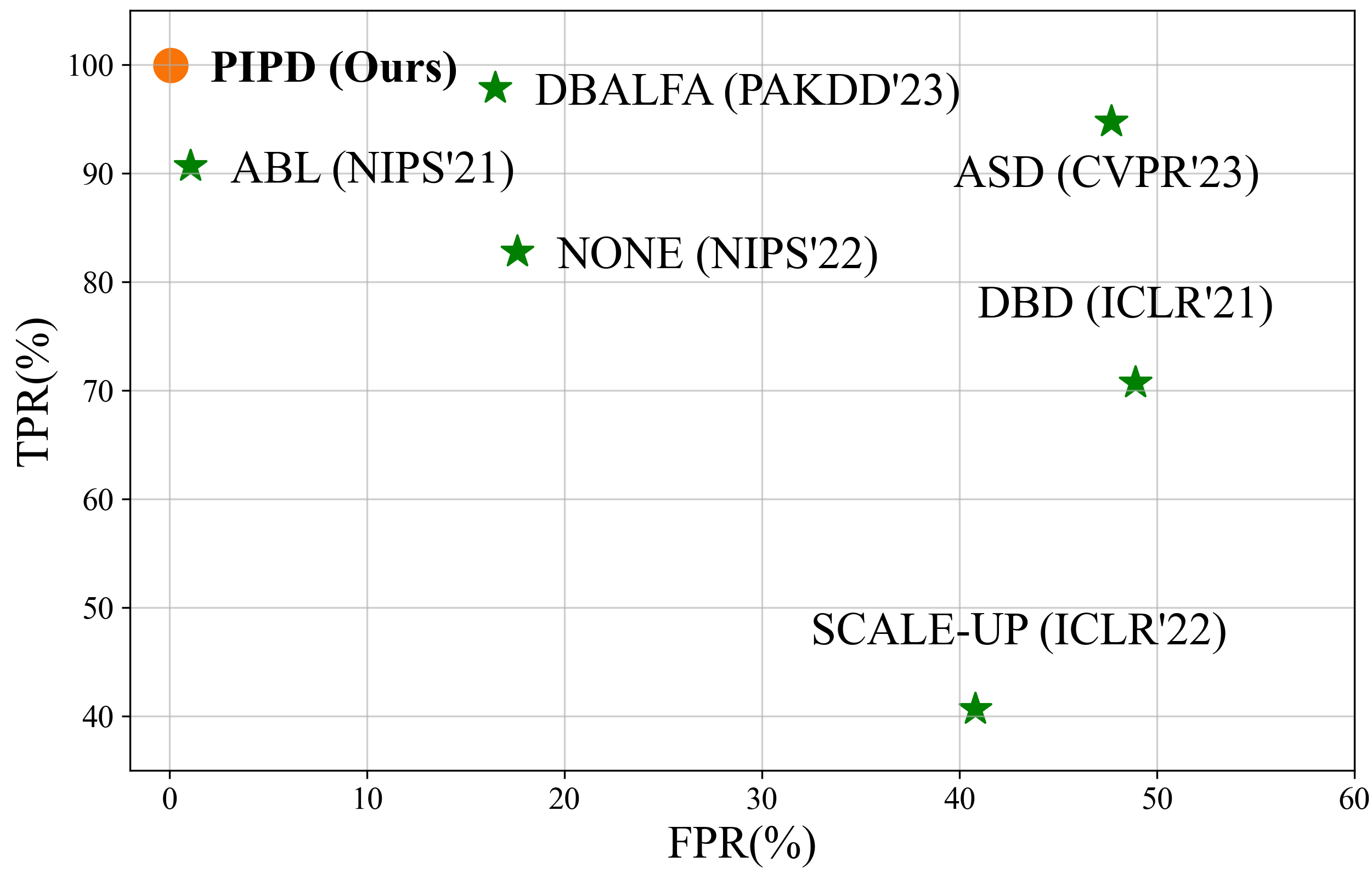}
\caption{Isolation quality comparison of the proposed PIPD with the state-of-the-art methods on CIFAR-10 \cite{cifar}.}
\label{SOTA}
\end{figure}

As backdoor attacks pose significant threats to DNNs, many strategies and techniques have been proposed to effectively defend against these attacks. This paper focuses on the training-time defense that involves isolating poisoned samples, thereby enabling the training of a clean model on a poisoned dataset. There are several representative works in this topic including \cite{ABL, DBD, NONE, ASD}. Specifically, \cite{ABL} leveraged the training loss to isolate poisoned samples then apply an unlearning process to train a clean model on poisoned dataset. \cite{DBD} maintained two dynamically updated poisoned and benign data pool based on sample losses then used semi-supervised learning to fine-tune the model on these pools. Similarly, \cite{ASD} applied loss-guided split and meta-learning-inspired split to dynamically update two data pools. \cite{NONE} identified poisoned sample by checking linearity of trained models. Most of these methods initially perform a one-time isolation process to identify poisoned samples, then train a clean model during a subsequent training phase.


In this work, our aim is to establish a defense method based on isolating poisoned samples that extends one-time isolation to a progressive isolation for better isolation quality. We start by pinpoint a prevalent problem in current solutions where an one-time isolation strategy struggles to accurately distinguish poisoned samples within the poisoned dataset, and regularly mis-identify the benign samples as poisoned ones. The inaccurate separation will precipitate a high Attack Success Rate (ASR) in the post-training model, whereas the mis-identification problem could risk undermining the clean inference performance of the model. To this end, we propose a new defense process in which we firstly employ a pre-isolation process to initialize the poisoned and benign subsets for the later isolation process which based on distribution discrepancy. This process fortifies isolation quality by stringently controlling the isolation ratio, isolating only the samples with extreme training losses as the subset. The next stage involves identifying the channels exhibiting the most significant discrepancies and discerning two distributions based on the cumulative feature value in these channels. We tackle the high benign sample mis-identification problem by developing a progressive isolation strategy. This progressive process transforms one-time isolation into a progressive isolation, enhancing the quality of isolation progressively. Upon identifying the poisoned samples, we develop a selective training scheme to inhibit the model from learning the backdoor behavior. Employing these innovative strategies allows us to train a benign model on poisoned dataset exhibiting a low ASR and superior Clean Accuracy (CA). Combining the above designs, PIPD can surpass the state-of-the-art methods by a big margin in terms of the isolation result (see Fig.~\ref{SOTA}).

Our major contributions can be summarized as follows:
\begin{itemize}
    \item We propose a novel and effective poisoned data isolation strategy against backdoor attack. E.g., for BadNets, Blended, and Trojan attacks on the CIFAR-10 dataset, our isolation quality can reach 100\% TPR and 0\% FPR.
    
    \item We design a selective training strategy to effectively prevent the model from being implanted with backdoor after the poisoned part is isolated. In conjunction with our isolation technique, the proposed learning strategy can efficaciously suppress the ASR.
    \item Experiments across different networks and datasets have proved the effectiveness of our defense method. For almost every attack methods, our method succeeds in precise poisoned sample isolation and the ASR of the purified model is close to that training on clean set.
\end{itemize}
\begin{figure*}[t]
\centering
\includegraphics[width=\textwidth]{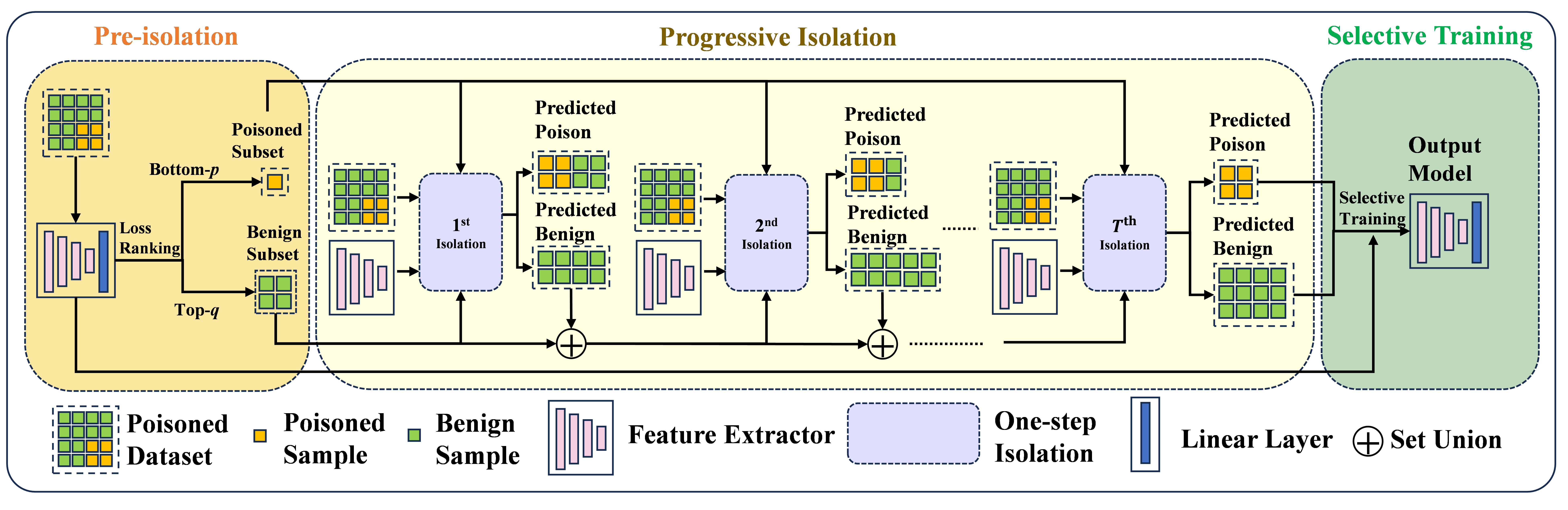}
\caption{The overview of PIPD method.}
\label{fig:framework}
\end{figure*}

\section{Related Works}

\subsection{Backdoor Attack}

As the pioneering work in backdoor attacks, \cite{BadNets} manipulated the training set by injecting mislabeled poisoned samples with a specific trigger to execute the attack. \cite{Trojan} selected an input region along with neurons which exhibit sensitivity to changes in that region, and the attacker aimed to activate these neurons. Many other works such as \cite{Blend, Dynamic} have proposed techniques to enhance the stealthiness of backdoor attacks, thereby reducing the effectiveness of backdoor detection mechanisms. Various forms of the triggers have been developed, including sinusoidal strips (SIG) \cite{SIG}, reflection (ReFool) \cite{Refool}, and warping-based (WaNet) \cite{WaNet}. In particular, WaNet used a smooth warping field to generate backdoor images with inconspicuous modifications. LIRA \cite{LIRA} alternated between trigger generation and backdoor injection to learn visually stealthy triggers. Adversarial Embedding \cite{9230390} sought to enhance the latent indistinguishability of backdoor attacks by minimizing the distance between the latent distributions of backdoor inputs and clean inputs through adversarial regularization. BppAttack \cite{BppAttack} combined image quantization and dithering to implant the trigger into the dataset. Combined with style transfer technique, \cite{Cheng_Liu_Ma_Zhang_2021} proposed an image style trigger attack. Similarly, \cite{Jiang_2023_CVPR} presented a backdoor attack method using color space shift. Backdoor attacks have been extended to various applications, including those outlined by \cite{Saha_2022_CVPR} for self-supervised learning \cite{BYOL}, \cite{li-etal-2023-multi-target} for pretrained models and \cite{ma2022dangerous} for object detection \cite{redmon2018yolov3}. Concurrently, certain backdoor attacks have been specifically designed for distinct network architectures, exemplified by \cite{Yuan_2023_CVPR} for vision transformers \cite{vit} and \cite{Chou_2023_CVPR} for the diffusion model \cite{DDPM}. A handful of studies, such as that by \cite{hayase2023fewshot}, have also explored more efficient methods to implant backdoors, with their work focusing on neural tangent kernels to significantly reduce the poison rate. Additionally, \cite{BadGPT} have explored the vulnerability of the latest ChatGPT to backdoor attacks.

\subsection{Backdoor Defense}
To combat backdoor attacks, many defense algorithms have been proposed to train clean models on poisoned datasets. \cite{ABL} observed that poisoned samples typically have a lower training loss compared to benign samples. They proposed a two-stage process: firstly, isolating a few samples with the lowest losses, and secondly, unlearning the backdoor on these isolated samples. \cite{DBD} was the pioneer in utilizing self-supervised learning and semi-supervised learning in backdoor defense. \cite{SCALEUP} found that poisoned samples demonstrated scaled prediction consistency when pixel values were amplified, and they identified poisoned samples by tracking the predictions of these scaled images. \cite{teco} revealed that poisoned models exhibit consistent performance on benign images under various image corruptions while performing divergently on poisoned images. \cite{layer_wise} observed that poisoned samples and benign samples exhibit significant differences at a crucial layer, and using the feature difference at this critical layer could aid in distinguishing poisoned samples. \cite{NONE} pinpointed backdoor neurons by investigating the linearity of the models, and performed a statistical test to identify poisoned samples.

\section{Preliminaries}

\subsection{Threat Model}
We consider the same threat model as in prior defense methods \cite{ABL, DBD, ASD, NONE}, which assumed the backdoor injection is performed at training stage and the attacker can only poison the dataset rather than control the training process. Formally, given a training set $\mathcal{D} = \{\mathcal{D}_b, \mathcal{D}_p\}$ including poisoned part $\mathcal{D}_p$=$\left\{(\hat{x}_i, \hat{y}_i)\right\}_{i=1}^{n_p}$ and benign part $\mathcal{D}_b$=$\left\{(x_i, y_i)\right\}_{i=1}^{n_b}$, the training process can be regarded as a dual-task learning on these two parts:
\begin{multline}
    \theta^* = \mathop{\arg\min}\limits_{\theta}(\mathbb{E}_{(x,y)\sim\mathcal{D}_b}\big[\ell(f(x),y)\big] + \\ 
\mathbb{E}_{(\hat{x},\hat{y})\sim\mathcal{D}_p}\big[\ell(f(\hat{x}),\hat{y})\big]),
\end{multline}
where $\theta$ represents the parameters of model $f$, and $\ell$ denotes the objective loss (\emph{e.g.}, cross-entropy loss).  

\begin{figure*}[t]
\centering
\includegraphics[width=0.935\textwidth]{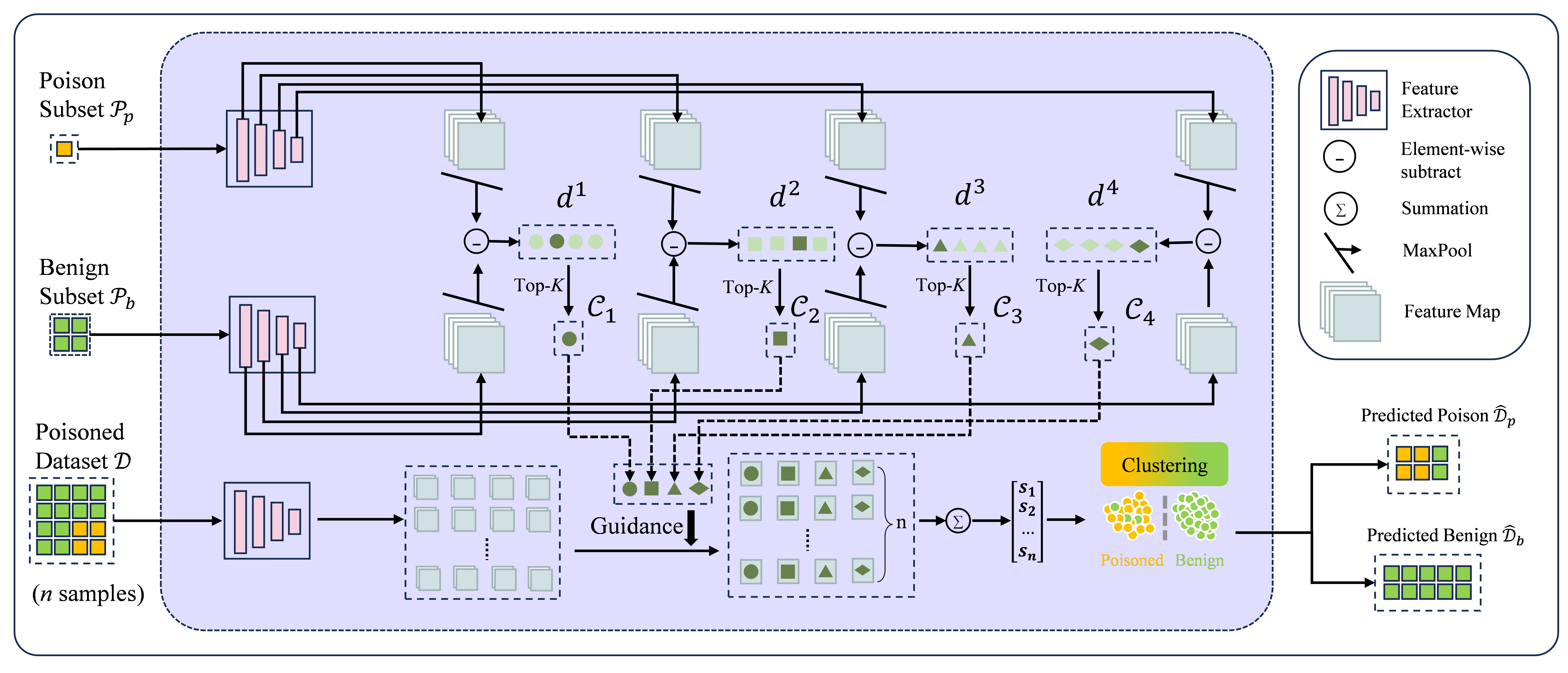}
\caption{One-step isolation process in PIPD.}
\label{fig:one-time-isolation}
\end{figure*}

\subsection{Defense Goal} 
Fundamentally, the objective of defense is to train a clean model on poisoned dataset $\mathcal{D}$. This model should have a low ASR on the poisoned test set, while maintaining a high inference performance on the clean test set. Considering the capabilities in previous works, we assume that the defender has access to the poisoned dataset $\mathcal{D}$ and is aware of the presence of poisoned data within it but lacks knowledge about the type of the attack or the poison rate $\eta$ which is the ratio of the number of poisoned samples to the total data size.

\section{Proposed PIPD Method}
\noindent\textbf{Motivation} Before delving into the specifics of our method, we illuminate two major deficiencies in existing methods: inferior isolation and high false alarm. Primarily, the utilization of one-time isolation of poisoned and benign samples usually results in unsatisfying isolation. For instance, ABL \cite{ABL}, for which the isolation is based on training loss, may fail to isolate poisoned samples since certain poisoned samples might still manifest high training loss, and different classes intrinsically have varying average losses. Secondly, existing methods tend to classify many benign samples as poisoned, consequently leading to an elevated FPR (see Fig.~\ref{SOTA}). This misidentification reduces the amount of data available for training and leads to an inferior clean inference performance. 

To overcome these aforementioned defects, we propose to extend the one-time isolation to a progressive isolation process, thereby refining isolation results through iterative progression and alleviating the aforementioned two challenges. Specifically, we design Progressive Isolation of Poisoned Data (PIPD) framework as illustrated in Fig.~\ref{fig:framework}. PIPD comprises two stages: the poisoned data isolation stage and the selective training stage. Within the stage of isolating poisoned data, there are further sub-divisions: the pre-isolation phase and the progressive isolation phase. The pre-isolation phase involves retrieving a compact, yet highly reliable poisoned subset and benign subset through a conventional loss-guided isolation method. Based on these subsets and in each iteration of isolation, we pinpoint the channels demonstrating the most significant differentiation, and subsequently cluster the dataset into poisoned and benign segments, considering the values of these channels. The predicted partitions are then harnessed for the subsequent iteration of isolation. Following the completion of the isolation process, the model is updated utilizing the predicted benign set, while a selective training process is applied to the identified poisoned set to hinder the model from incorporating the backdoor. Algorithm of our PIPD is shown in Appendix \ref{ap:algorithm}.

PIPD achieves isolation results on CIFAR-10 with an average TPR and FPR of 99.95\% and 0.06\% respectively, outperforming the competitors by margins of 2.12\% and 1.01\%. These superior isolation results provide strong support for our method to achieve low ASR and high CA for the model.

\subsection{Pre-isolation Phase}

We propose to utilize the isolation results of an existing isolation method as a setup, furnishing a precise prior for the subsequent isolation based on distributional discrepancies. By adopting the loss function from \cite{ABL}:
\begin{equation}
\label{eq:ABL_loss_eq}
    \mathcal{L}_{LGA} = \mathbb{E}_{(x,y)\sim\mathcal{D}}\big[\mathrm{sign}(\ell(f(x),y)-\gamma)\big]\cdot \ell(f(x),y),
\end{equation}
the samples with a slower loss descent rate will be trapped around a threshold $\gamma$, while keeping other samples away. After training for a few epochs, we select $\emph{p}$ lowest loss samples as poisoned subset $\mathcal{P}_p$ and $\emph{q}$ highest samples as benign subset $\mathcal{P}_b$. Considering the low poison rate $\eta$ (usually $\eta < 20\%$), we set $\emph{p}$ as 1\% and $\emph{q}$ as 20\%.

\subsection{Progressive Isolation Phase}
\subsubsection{One-step Isolation}

The objective of one-step isolation is to partition dataset $\mathcal{D}$ into $\hat{\mathcal{D}}_b$ and $\hat{\mathcal{D}}_p$, based upon the provided poisoned subset $\mathcal{P}_p$ and benign subset $\mathcal{P}_b$. This process $\mathcal{ISO}$ can be formulated as: 
\begin{equation}\label{eq:one-step-isolation}
    \{\hat{\mathcal{D}}_p, \hat{\mathcal{D}}_b\} = \mathcal{ISO}\Big(\mathcal{P}_b, \mathcal{P}_p, \mathcal{D}\Big).
\end{equation}
Inspired by the fact that for a poisoned model, any input containing a trigger will be predicted as the target label. This indicates that the neurons activated by the trigger dominate in the feature space. Consequently, the activation of poisoned samples exhibits a pronounced value difference compared to the activation of benign samples. Therefore, we leverage the previously obtained $\mathcal{P}_p$ and $\mathcal{P}_b$ to pinpoint the channels where the discrepancy between the poisoned and clean data distributions is at its zenith. Depending on these channels, we evaluate the entire dataset $\mathcal{D}$ and segregate the poisoned data. As shown in Fig.~\ref{fig:one-time-isolation}, we first retrieve the intermediate features for poisoned and benign subsets. To identify the channels that exhibit the maximum distribution discrepancy in each layer, a measurement of discrepancy $d_i^{l}$ of $i$-th channel in $l$-th layer is proposed:
\begin{equation}\label{eq:discrepancy_measure}
    d_i^{l} = \mathop{\max}\limits_{x_j\in\mathcal{P}_p}{f_i^l(x_j)} - (\mu_{b,i}^{l} + \beta \cdot \sigma_{b,i}^{l}),
\end{equation}
where
\begin{equation}\begin{aligned}
    \mu_{b,i}^{l} &= \frac{1}{|\mathcal{P}_b|} \sum_{x_j\in\mathcal{P}_b}f_{i}^{l}(x_j),\\
    \sigma_{b,i}^{l} &= \sqrt{\frac{1}{|\mathcal{P}_b|}\sum_{x_j\in\mathcal{P}_b}(f_{i}^{l}(x_j)-\mu_{b,i}^{l})}.\\
\end{aligned}
\end{equation}
Here, $f_{j}^{l}(x_i)$ represents the feature of sample $x_i$ in the $j$-th channel from \emph{l}-th layer and $\beta$ is empirically set as 3 for balancing the mean and standard deviation. We calculate the maximum value for poisoned subset (first term in Eq. \eqref{eq:discrepancy_measure}) rather than mean and standard deviation as benign subset for the ensuing reason: the size of the poisoned subset is diminutive, and the inadvertent inclusion of benign samples can easily perturb the mean and standard deviation of its distribution. Section \ref{ab:fault} provides experiments to demonstrate the stability of our design against such faults in the poisoned subset. We identify the distinct channels $\mathcal{C}_l$ in the \emph{l}-th layer as the $K$ channels with the largest discrepancy.
\begin{equation}
\label{distribution_discrepancy_eq}
    \mathcal{C}_l = \mathop{TopK}\limits_{i\in\{1...c_l\}}d_i^{l}.
\end{equation}
Next, we aggregating the feature values from the distinct channel for each sample:
\begin{equation}\label{eq:scores}
    \mathcal{S} = \Big\{s_i = \sum_{l=1}^{L}\sum_{j\in\mathcal{C}_l}f_{j}^{l}(x_i)\mid x_i \in \mathcal{D}\Big\}_{i=1}^{n}.
\end{equation}
The last step entails utilizing a clustering method to segregate the poisoned data from the entire dataset, employing the scores previously calculated. Here, we adopt the Fisher-Jenks algorithm for its simplicity:
\begin{equation}\label{eq:cluster}
    \{\hat{\mathcal{D}}_p, \hat{\mathcal{D}}_b\} = cluster(\mathcal{S}),
\end{equation}
where $cluster(\cdot)$ represents the clustering algorithm.

\subsubsection{Progressive Isolation}
We now elaborate how one-step isolation can be extended to a progressive isolation process. As is depicted in the progressive part of the Fig.~\ref{fig:framework}, the progressive isolation comprises $T$ iterations of one-step isolation. In each iteration, there are four inputs, including $\mathcal{P}_b$, $\mathcal{P}_p$, $\mathcal{D}$ and the feature extractor. Apart from the first iteration where $\mathcal{P}_b$ is retrieved from the pre-isolation phase, in all subsequent iterations, $\mathcal{P}_b$ is composed of the union of $\hat{\mathcal{D}}_b$ outputted from the previous iteration and $\mathcal{P}_b$ inputted to that previous iteration. In contrast, $\mathcal{P}_p$ consistently remains unchanged. The $t$-th one-step isolation finishes with two outputs including $\hat{\mathcal{D}}_b^t$ and $\hat{\mathcal{D}}_p^t$. After the maximum allowable iteration number is reached, the final $\hat{\mathcal{D}}_b^T$ and $\hat{\mathcal{D}}_p^T$ will serve for the next stage of training. The progressive process at the $t$-th epoch can be defined as:
\begin{equation}\label{eq:enlarge}
\begin{aligned}
    \{\hat{\mathcal{D}}_b^{t+1}, \hat{\mathcal{D}}_p^{t+1}\} &= \mathcal{ISO}\Big(\mathcal{P}_b^{t}, \mathcal{P}_p, \mathcal{D}\Big), \\
    \mathcal{P}_b^{t+1} &= \hat{\mathcal{D}}_b^{t+1} \cup \mathcal{P}_b^{t}.
\end{aligned}
\end{equation}

In the case of one-step isolation, the elevated FPR issue can be ascribed to the fact that certain benign samples and poisoned samples remain indistinguishable on these channels. This implies that the identified channels might be imprecise. In Eq. \eqref{eq:discrepancy_measure}, we calculate the mean and standard deviation of the feature values within the benign subset, offering a statistical perspective of the benign distribution. Empirically, enlarging the number of samples enhances the precision of this statistical portrayal of the benign distribution, thereby aiding in the identification of more apt channels where a greater number of benign samples and poisoned samples can be separated.

\subsection{Selective Training Phase}
After acquiring the final predicted poisoned set $\hat{\mathcal{D}}_p^{T}$ and benign set $\hat{\mathcal{D}}_b^{T}$, we propose to optimize the $\theta$ of $f$ via a selective training strategy rather than conventional training. Specifically, we only execute gradient descent when $f$ identifies the samples within $\hat{\mathcal{D}}_p^{T}$ as their ground-truth labels, while employ standard gradient descent on $\hat{\mathcal{D}}_b^{T}$. The training process can be formulated as:
\begin{multline}
\label{eq:unlearn_schema_1}
\theta^{*} = \mathop{\arg\min}\limits_{\theta}(\mathbb{E}_{(x,y)\sim\hat{\mathcal{D}}_b^{T}}\big[\ell(f(x),y)\big] - \\ 
\lambda \cdot \mathbb{E}_{(\hat{x},\hat{y})\sim\hat{\mathcal{D}}_p^{T}}\big[\mathds{1} (f(\hat{x}) = \hat{y})\ell(f(\hat{x}),\hat{y})\big]),
\end{multline}
where the indicator function $\mathds{1}(\cdot)$ returns 1 when the condition is true and 0 otherwise. Compared with a similar solution in ABL, in which it continues to apply gradient ascent on the isolated samples, our strategy is capable of mitigating the impact of gradient ascent on the inference performance. 

\section{Experimental Results}

\begin{table*}[tp!]
\renewcommand{\arraystretch}{1.05} 
\centering
\begin{adjustbox}{width=0.98\linewidth}
\begin{tabular}{c|c|rr|rr|rr|rr|rr|rr}
\toprule
\addlinespace[0pt]
\multirow{2}{*}{Datasets} & \multirow{2}{*}{\begin{tabular}[c]{@{}c@{}}Backdoor \\ Attacks\end{tabular}} & \multicolumn{2}{c|}{No Defense} & \multicolumn{2}{c|}{ABL} & \multicolumn{2}{c|}{DBD} & \multicolumn{2}{c|}{NONE} & \multicolumn{2}{c|}{ASD} & \multicolumn{2}{c}{PIPD (Ours)} \\ \cline{3-14} 
 &  & ASR & CA & ASR & CA & ASR & CA & ASR & CA & ASR & CA & ASR & CA \\ \hline
\multirow{10}{*}{CIFAR-10} & BadNets & 100.00 & 93.58 & 1.11 & 92.48 & 0.96 & 92.41  & 1.97 & 92.38 & 1.26 & 93.45 & \textbf{0.73} & \textbf{94.63} \\
 & Trojan & 100.00 & 93.53 & 1.97 & 92.46 & 8.02 & 92.17 & 24.93 & 93.95 & 0.91 & 93.79 & \textbf{0.43} & \textbf{94.38} \\
 & Blended & 100.00 & 94.00 & 1.65 & 91.90 &  1.73 & 92.18 & 99.57 & \textbf{94.27} & 1.49 & 92.90 & \textbf{0.03} & 94.12 \\
 & CL & 100.00 & 94.84 & 1.32 & 87.60 & 0.13 & 90.60 & 6.33 & 94.12 & 0.93 & 93.10 & \textbf{0.02} & \textbf{94.80} \\
 & SIG & 94.85 & 93.62 & 4.82 & 91.40 & 6.15 & 90.14 & 47.46 & 94.03 & 0.58 & 92.79 & \textbf{0.21} & \textbf{94.65} \\
 & Dynamic & 99.98 & 93.51 & 5.16 & 93.45 & 8.48 & 92.12 & 9.43 & \textbf{94.17} & 1.37 & 93.26 & \textbf{0.23} & 93.80 \\
 & WaNet & 99.10 & 93.67 & 2.23 & 84.15 & 0.39 & 91.20 & 94.86 & 93.18 & 2.31 & 92.28 & \textbf{0.91} & \textbf{94.48}  \\
 & A-Patch & 96.46 & 94.76 & 4.77 & 89.10 & 5.13 &  90.79  & 21.23 & 94.19 & 5.37 & 91.97 & \textbf{1.19} & \textbf{94.86} \\
 & Refool & 99.85 & 94.81 & 1.32 & 82.17 &  0.56 & 91.50 & 64.27 & 93.14 & 0.74 & 93.57 &\textbf{0.08} & \textbf{94.79}
\\ 
 \cline{2-14} 
 & Average & 98.91 & 94.04 & 2.70 & 89.41 & 3.51 & 91.45  & 41.11 & 93.71 & 1.66 & 93.01 & \textbf{0.43} & \textbf{94.50} \\ \hline
\multirow{5}{*}{ImageNet} & BadNets-Grid & 100.00 & 85.93 & 3.74 &  84.76  & 1.61 & 83.79 & 0.45 & 84.01 & 1.26 & 84.73 & \textbf{0.24} & \textbf{85.69} \\
 & Trojan-WM & 100.00 & 85.87 & 3.43 & 84.80 & 2.39 & 84.02 & \textbf{0.00} & 84.48 & 0.06 & 83.58 & 0.20 & \textbf{85.60} \\
 & Blended & 99.80 & 86.67 & 21.41 & 85.12 & 2.44 & 83.36 & 8.53 & 82.46 &  6.67 & 84.64 &  \textbf{0.00} & \textbf{86.13} \\
 & SIG & 43.53 & 86.67 & 6.76 & 81.10 & 4.52 & 81.59 & 49.57 & 81.07 &  5.78 & 85.85 &  \textbf{0.13} & \textbf{86.07} \\
 \cline{2-14} 
 & Average & 85.83 & 86.28 & 8.83 & 83.95 & 2.74 & 83.19 &  14.63  & 83.00 & 3.44 & 84.70 &  \textbf{0.14} & \textbf{85.87} \\ \hline
\end{tabular}
\end{adjustbox}
\caption{Performances of our method along with 4 competing backdoor defense methods against 9 backdoor attacks. The experiments are conducted over CIFAR-10 and ImageNet with ResNet-18. The best results are \textbf{boldfaced}.}
\label{tab:model_performance}
\end{table*}

\begin{table*}[t]
\centering
\begin{adjustbox}{width=0.98\linewidth}
\begin{tabular}{c|rrrrrrrrrrrrrr|rr}
\toprule
\addlinespace[0pt]
\multicolumn{1}{c|}{Attack$\rightarrow$}         & \multicolumn{2}{c}{BadNets}        & \multicolumn{2}{c}{Blended}     & \multicolumn{2}{c}{SIG}          & \multicolumn{2}{c}{Dynamic}      & \multicolumn{2}{c}{Trojan}      & \multicolumn{2}{c}{CL}  & \multicolumn{2}{c|}{Refool}     & \multicolumn{2}{c}{Avg}        \\ \hline
\multicolumn{1}{c|}{\tabincell{c}{Metric$\rightarrow$\\Defense$\downarrow$}} & TPR          & FPR         & TPR           & FPR        & TPR          & FPR         & TPR         & FPR        & TPR         & FPR         & TPR          & FPR & TPR          & FPR         & TPR          & FPR         \\ \hline
ABL & 94.76 & 0.68 & 87.68 & 0.64 & 90.60 & 0.49 & 93.00 & 1.42 & 88.44 & 2.18 & 92.52 & 1.44 & 87.48 & 0.65 & 90.64 &  1.07 \\
DBD & 80.67 & 48.52 & 99.44 & 47.39 & 38.36 & 50.61 & 21.48 & 51.50 & 99.64 & 47.38 & 58.72 & 49.54 & 96.6 & 47.54 & 70.70 & 48.92 \\
NONE & 99.80 & 18.75 & 78.84 & 33.28 & 21.08 & 36.38 & 99.84 & 0.01 & \textbf{100.00} & 0.74 & 99.90 & 0.11 & 79.92 & 34.07 & 82.76 & 17.62 \\
DBALFA & 99.79 & 7.19 & 97.54 & 7.30 & 95.30 & 30.35 & 98.28 & 16.55 & 99.25 & 15.84 & 98.66 & 7.23 & 96.02 & 31.06 & 97.83 & 16.50\\ 
SCALE-UP & 32.36 & 32.54 & 39.36 & 40.05 & 27.64 & 27.38 & 28.72 & 27.79 & 30.56 & 30.98 & 38.84 & 40.12 & 86.96 & 86.83 & 40.63 & 40.81 \\
ASD & 97.77 & 47.70 & 100.00 & 47.36 & 99.30 & 47.63 & 97.92 & 47.47 & 96.36 & 47.56 & 86.92 & 48.05 & 84.92 & 48.16 & 94.74 & 47.70 \\ \hline
PIPD (Ours) & \textbf{100.00} & \textbf{0.00} & \textbf{100.00} & \textbf{0.00} & \textbf{100.00} & \textbf{0.00} & \textbf{100.00} & \textbf{0.00} & \textbf{100.00} & \textbf{0.00} & \textbf{100.00} & \textbf{0.00} & \textbf{99.65} & \textbf{0.42} & \textbf{99.95} & \textbf{0.06} \\ \addlinespace[-0.22em]
\bottomrule[0.68pt]
\end{tabular}
\end{adjustbox}
\caption{TPR/FPR isolation results (\%) on the CIFAR-10 dataset. The best results are \textbf{boldfaced}.}
\label{tab:isolation}
\end{table*}

\subsection{Experimental Setup}
\textbf{Dataset $\mathcal{D}$}: Following convention \cite{ABL, NONE, DBD, ASD}, we conduct experiments over the CIFAR-10 \cite{cifar} and a subset of ImageNet \cite{ImageNet} datasets. 

\noindent\textbf{Attack Setups}: We choose nine state-of-the-art backdoor attacks, including BadNets \cite{BadNets}, Trojan \cite{Trojan}, Blended \cite{Blend}, Dynamic \cite{Dynamic}, WaNet \cite{WaNet}, SIG \cite{SIG}, CL \cite{CL}, A-Patch \cite{qi2023revisiting}, and Refool \cite{Refool}. Unless otherwise stated, the target label is designated as class 0. The poison rate is set to 5\% for all attacks, with the exception of those categorized as clean-label attacks (\emph{e.g.}, SIG and CL).

\noindent\textbf{Competing Methods}: We compare our method on model performance with four training-time defense methods, including ABL \cite{ABL}, DBD \cite{DBD}, NONE \cite{NONE} and ASD \cite{ASD}. As for the isolation quality comparison, we additionally include two detection methods DBALFA \cite{layer_wise} and SCALE-UP \cite{SCALEUP}. Note that for SCALE-UP, We select the best threshold with the highest TPR and the lowest FPR.

\noindent\textbf{Evaluation Metrics}: We evaluate the performance by using the following metrics, namely, CA, ASR, TPR, and FPR. High CA and TPR, low ASR and FPR are desired.

\noindent\textbf{Implementation Details}: We employ ResNet-18 \cite{he2016deep} as our default network. During the one-step isolation process, we extract the feature maps subsequent to each convolutional layer. The pre-isolation epoch is designated at 200, with the progressive iteration number $T$ set to 8, and the epochs for selective training is 20.

Due to space limit, more details regarding the experiments are deferred to the appendix. Specifically, the isolation results of PIPD on poisoned dataset with different poison rates, and the impact of clustering method on isolation results are included in Appendix \ref{ap:poison-rate} and \ref{ap:cluster}, respectively.

\subsection{Comparisons on Defense Performance}

The ASR and CA results are delineated in Table.~\ref{tab:model_performance}. As the data illustrates, our defense method consistently achieves superior CA and low ASR values in all attack settings. Specifically, PIPD demonstrates the highest average CA at 94.5\% and the lowest average ASR at 0.43\%. For the more complex ImageNet, PIPD still exhibit superior performance, where the average ASR is 0.14\% and the average CA is 85.87\%.

As a comparision, we also present the defense performances of other defense methods. Both the DBD and ASD methods employ semi-supervised learning, and they exhibit relatively low ASRs. This is attributed to their treatment of a significant portion of data as unlabeled, whereby the removal of labels from poisoned data effectively prevents backdoor injections. The CA for the NONE method is the highest among them, and simultaneously, its ASR is also the highest. It's noteworthy that the NONE method employs two mechanisms to reduce ASR: one through isolating samples and another through resetting neurons. A high CA suggests that it isolates an insufficient number of samples, leaving ample residual data for training. However, the elevated ASR indicates that the neurons it resets are not necessarily related to the backdoor. The CA results for the ABL method are suboptimal. We attribute this to the detrimental effects of excessive unlearning on the model's performance. Benefiting from the progressive isolation process with selective training strategy, our PIPD leads to the best performance over all testing datasets in both CA and ASR. These results confirm the effectiveness of our proposed defense method in training a clean model using a poisoned dataset.

\subsection{Comparisons on Isolation Quality}
We now elucidate the isolation quality of our algorithm and juxtapose it with established isolation methods. Table.~\ref{tab:isolation} reports the isolation quality. Our method demonstrates the capacity to yield the highest TPR and lowest FPR across all attacks. Notably, apart from the Refool method, the TPR values with other attacks reach at 100\%, signifying that most of the poisoned samples are isolated. Also, the FPR for all attacks are both 0\%, except from a 0.42\% for the Refool. It can be seen that, by using the proposed PIPD, benign samples are rarely classified as poisoned. In comparison, ABL, which records the second lowest FPR of 1.07\% among these methods, is still 1.01\% higher than PIPD. Concurrently, the average TPR value of ABL lags ours by 9.31\%. It should be noted that DBALFA requires a 10\% benign test set for setup, yet its TPR falls short of ours by 2.12\%, even though it exhibits the highest TPR amongst other methods. The FPR values of DBD and ASD are comparably high, which is the consequence of treating 50\% of the dataset as poisoned. These findings solidify our method's capability to deliver low ASR and high CA for the model.

\section{Ablation study}
We now analyze how each component contributes to the PIPD in terms of pre-isolation process, progressive isolation process and selective training strategy. Unless otherwise stated, the network being employed is ResNet-18 and the dataset is CIFAR-10 with a 5\% poison rate.

\begin{table}[!tbp]
\footnotesize
\setlength\tabcolsep{3pt}
\scalebox{0.88}{
\begin{tabular}{l|rrrrrrrr}
\toprule[0.68pt]
\addlinespace[0pt]
\multicolumn{1}{c|}{Fault Ratio$\rightarrow$} & \multicolumn{2}{c}{20\%} & \multicolumn{2}{c}{40\%} & \multicolumn{2}{c}{60\%} & \multicolumn{2}{c}{80\%} \\ \cline{1-9} 
\multicolumn{1}{c|}{Attack$\downarrow$} & TPR & FPR & TPR & FPR & TPR & FPR & TPR & FPR\\ \hline
BadNets & 100.00 & 0.00	& 100.00 & 0.00 & 100.00 & 0.00 & 100.00 & 0.00\\
Blended & 100.00 & 0.00	& 100.00 & 0.00 & 100.00 & 0.00 & 100.00 & 0.00\\
Dynamic & 99.96 & 0.00	& 99.92 & 0.00 & 99.92 & 0.00 & 99.92 & 0.00\\ 
Trojan & 100.00 & 0.00	& 100.00 & 0.01 & 100.00 & 0.02 & 100.00 & 0.62\\ 
A-Patch & 100.00 & 0.00	& 99.76 & 0.06 & 99.68 & 0.06 & 99.72 & 0.10\\
\addlinespace[-0.22em]
\bottomrule[0.68pt]
\end{tabular}}
\caption{TPR/FPR results (\%) of our PIPD when poisoned subset contains benign samples.}
\label{tab:fault_tolerance}
\end{table}

\subsection{The Impact of Pre-isolation}
\label{ab:fault}
In this experiment, we explore the scenario where the poisoned subset, derived from the pre-isolation phase, encompasses a number of benign samples. The proportion of the poisoned subset is 1\%, assembled by randomly selecting diverse ratios of poisoned to benign samples. We define the fault ratio as the percentage of benign samples mixed in the poisoned subset. Specifically, we scrutinize settings where the fault ratio stands at 20\%, 40\%, 60\%, and 80\%. Table.~\ref{tab:fault_tolerance} presents the isolation performance of our PIPD approach. It has been observed that as more benign samples are mixed into the poisoned subset, the TPR displays a downward trend, while the FPR shows an upward trajectory. This implies that an inferior quality of the poisoned subset will compromise the final isolation quality.

\subsection{The Impact of Progressive Isolation}
\begin{figure}
    \centering 
    \includegraphics[width=0.84\columnwidth]{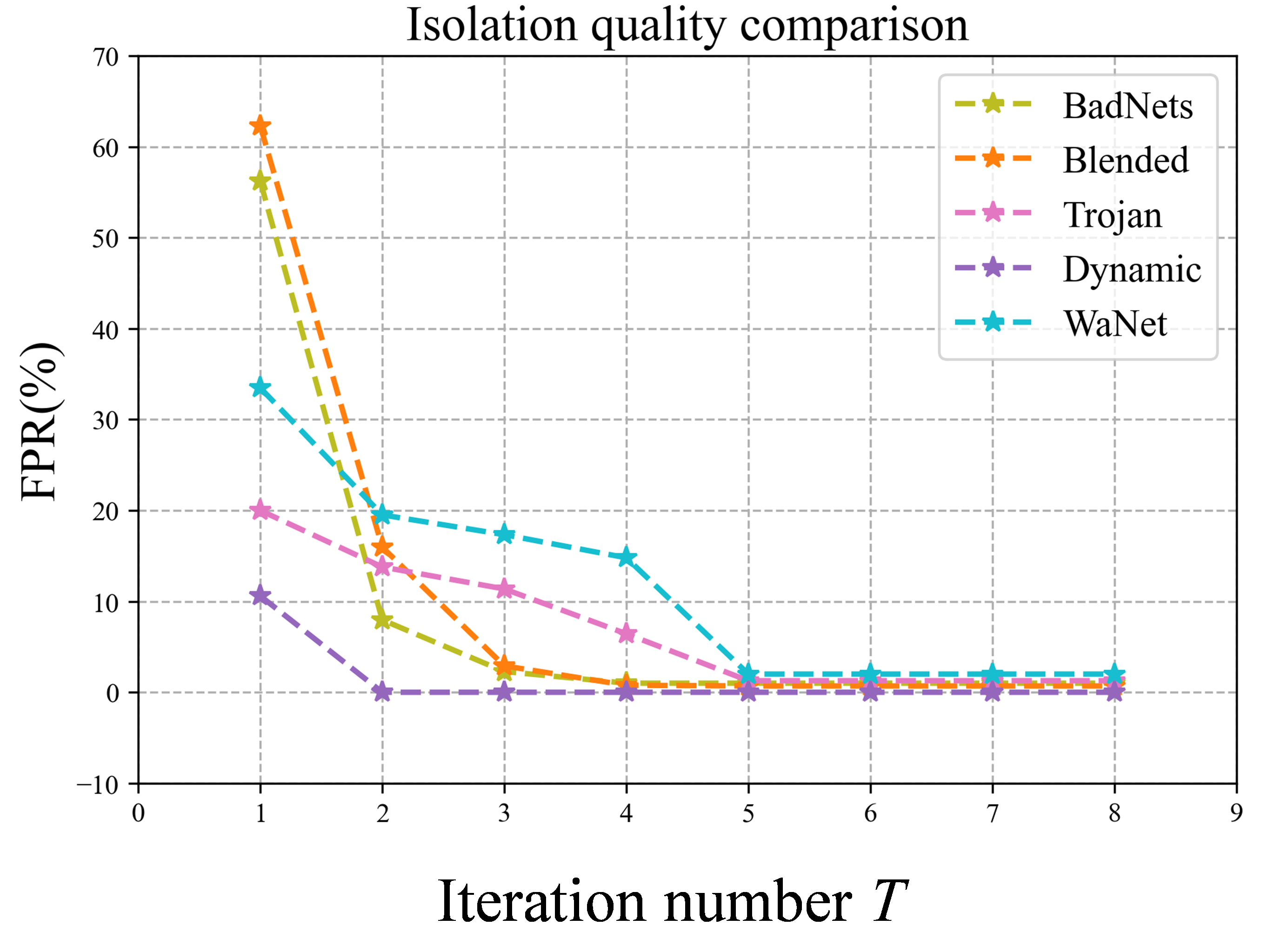}
    \caption{FPR values of PIPD for different $T$.}
    \label{Converge evidence}
\end{figure}

This section aims to testify if the progressive isolation process effectively mitigates the issue of high FPR. In this experiment, we choose a poison rate of 20\% for illustration. As demonstrated in Fig.~\ref{Converge evidence}, the preliminary isolation process yields suboptimal results, specifically an average FPR value of around 0.3 with these attacks. However, with the progressive refinement of the isolation results, the FPR decreases constantly. These outcomes strongly endorse the effectiveness of our progressive strategy in addressing the high FPR issue. Additionally, we evaluate the CA and ASR under different $T$s, and the results are displayed in Table.~\ref{tab:different_progressive_epoch}. The isolation process of the first iteration failed to defense against BadNet and Blended with a high ASR of 99.81\% and 98.73\%. By gradually increasing $T$, the ASR values decline constantly. Meanwhile, the CA values increase when $T$ becomes larger, indicating that the number of benign samples being misclassified is decreasing and the FPR is effectively reduced. We conjecture that five iterations is enough for defending most attacks; while increasing $T$ merely prolongs the isolation time.

\begin{table}[t]
\footnotesize
\centering
\setlength{\tabcolsep}{0.9mm}{
\scalebox{1.0}{
\begin{tabular}{l|rrrrrrrr}
\toprule[0.68pt]
\addlinespace[0pt]
\multicolumn{1}{c|}{\multirow{2}{*}{$T$}} & \multicolumn{2}{c}{BadNets} & \multicolumn{2}{c}{Blended} & \multicolumn{2}{c}{Trojan} & \multicolumn{2}{c}{Dynamic} \\ \cline{2-9} 
\multicolumn{1}{c|}{} & CA & ASR & CA & ASR & CA & ASR & CA & ASR\\ \hline
1 & 92.98 & 99.81 & 93.28 & 98.73 & 90.80 & 56.22 & 93.73 & 1.65 \\
2 & 94.14 & 0.54 & 93.61 & 50.46 & 94.09 & 13.75 & 94.27 & 1.45 \\
3 & 94.24 & 0.54 & 94.06 & 0.72 & 94.04 & 1.94 & 94.15 & 1.41 \\
4 & 94.22 & 0.51 & 94.01 & 0.69 & 94.08 & 1.87 & 94.22 & 1.42 \\
5 & 94.23 & 0.52 & 94.04 & 0.72 & 94.09 & 1.63 & 94.21 & 1.46 \\
\addlinespace[-0.22em]
\bottomrule[0.68pt]
\end{tabular}}}
\caption{CA/ASR(\%) of our PIPD with different $T$.}
\label{tab:different_progressive_epoch}
\end{table}

\subsection{The Impact of Selective Training}
\begin{figure}
    \centering 
    \includegraphics[width=0.87\columnwidth]{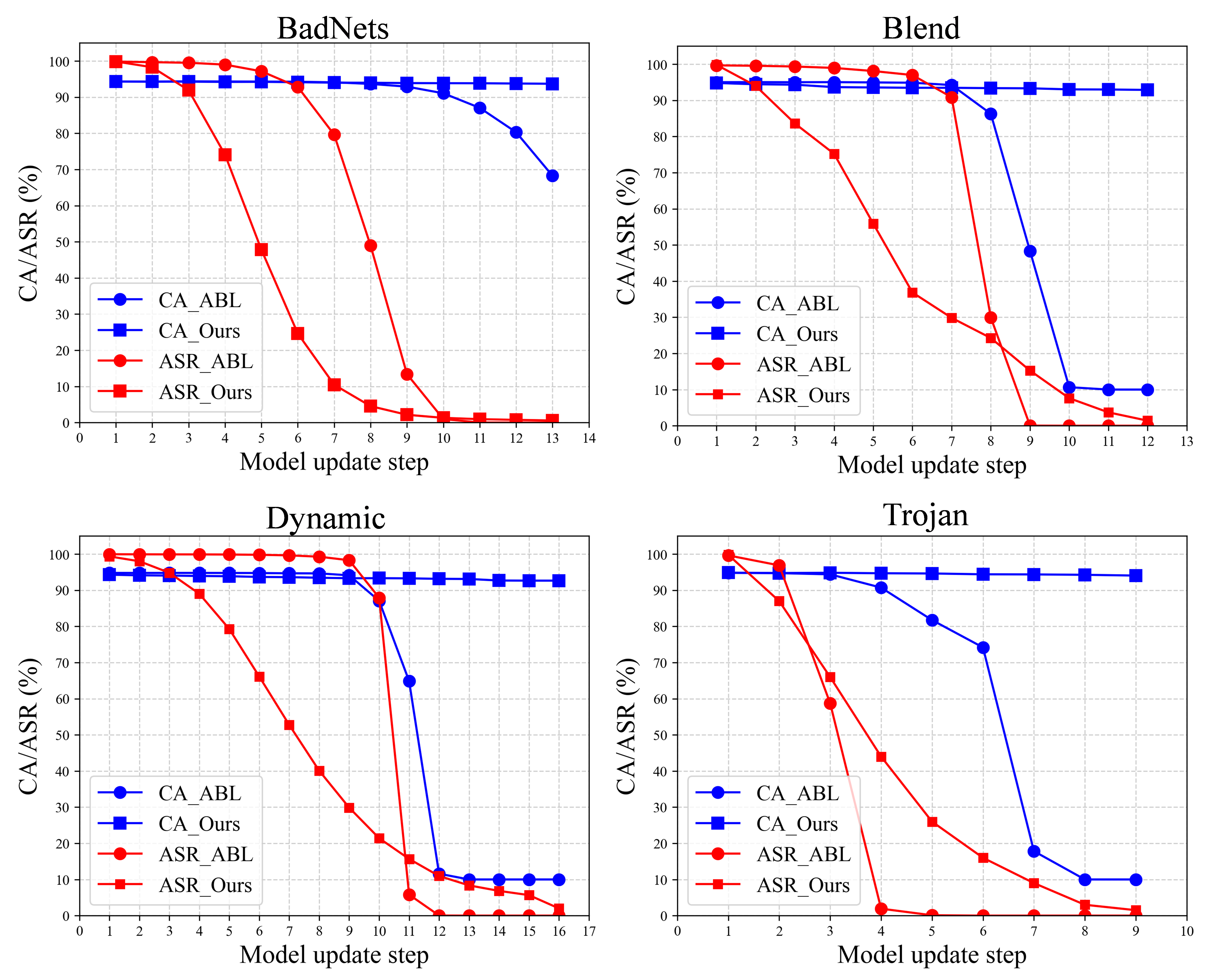}
    \caption{The defense result comparison between unlearning process from ABL and selective training in PIPD.}
    \label{tab:selective-trianing-ablation}
\end{figure}

This section studies the effect on defense performance with selective training strategy, especially compared with the unlearning strategy in ABL. For fair comparison, we use the same isolation results and only replace the training strategy. As can be seen from Fig.~\ref{tab:selective-trianing-ablation}, our approach consistently maintains a high CA, during the training process as the ASR declines. In contrast, regarding the ABL method, as the ASR decreases, the CA also decreases sharply. We ascribe this to excessive unlearning damaging the inference performance of the model on clean test set. In our method, the model does not perform unlearning on poisoned samples that are no longer predicted as the target label. This averts the detrimental effects of excessive unlearning on the CA performance.

\section{Conclusion}

In this paper, we introduce a novel training-time backdoor defense method, PIPD, premised on a progressive data isolation process. We extended the conventional one-time isolation approach to a progressive isolation process, yielding improved isolation results in terms of TPR and FPR. By combining this with a new selective training strategy, we effectively train a clean model on a poisoned dataset, . The robustness and effectiveness of our PIPD are validated through extensive experiments. 

Despite its merits, PIPD does have limitation. Specifically, it cannot counter attacks that necessitate dynamically adjusting poisoned samples, especially where the attacker controls the entire training process. Additionally, the extension of our PIPD method to other challenges, such as label noise, will be explored as future work.

\newpage

\section*{Acknowledgments}
This work was supported in part by Macau Science and Technology Development Fund under SKLIOTSC-2021-2023, 0072/2020/AMJ and 0022/2022/A1; in part by Research Committee at University of Macau under MYRG2022-00152-FST and  MYRG-GRG2023-00058-FST-UMDF; in part by Natural Science Foundation of China under 61971476; and in part by Alibaba Group through Alibaba Innovative Research Program.

\bibliography{aaai24}

\begin{thebibliography}{39}
\providecommand{\natexlab}[1]{#1}

\bibitem[{Barni, Kallas, and Tondi(2019)}]{SIG}
Barni, M.; Kallas, K.; and Tondi, B. 2019.
\newblock A New Backdoor Attack in CNNS by Training Set Corruption Without
  Label Poisoning.
\newblock In \emph{IEEE International Conference on Image Processing}.

\bibitem[{Chen et~al.(2017)Chen, Liu, Li, Lu, , and Song}]{Blend}
Chen, X.; Liu, C.; Li, B.; Lu, K.; ; and Song, D. 2017.
\newblock Targeted Backdoor Attacks on Deep Learning Systems Using Data
  Poisoning.
\newblock \emph{arXiv preprint arXiv:1712.05526}.

\bibitem[{Cheng et~al.(2021)Cheng, Liu, Ma, and
  Zhang}]{Cheng_Liu_Ma_Zhang_2021}
Cheng, S.; Liu, Y.; Ma, S.; and Zhang, X. 2021.
\newblock Deep Feature Space Trojan Attack of Neural Networks by Controlled
  Detoxification.
\newblock \emph{Proceedings of the AAAI Conference on Artificial Intelligence}.

\bibitem[{Cheng et~al.(2023)Cheng, Tao, Liu, An, Xu, Feng, Shen, Zhang, Xu, Ma,
  and Zhang}]{beagle}
Cheng, S.; Tao, G.; Liu, Y.; An, S.; Xu, X.; Feng, S.; Shen, G.; Zhang, K.; Xu,
  Q.; Ma, S.; and Zhang, X. 2023.
\newblock BEAGLE: Forensics of Deep Learning Backdoor Attack for Better
  Defense.
\newblock \emph{Network and Distributed System Security Symposium}.

\bibitem[{Chou, Chen, and Ho(2023)}]{Chou_2023_CVPR}
Chou, S.-Y.; Chen, P.-Y.; and Ho, T.-Y. 2023.
\newblock How to Backdoor Diffusion Models?
\newblock In \emph{Proceedings of the IEEE/CVF Conference on Computer Vision
  and Pattern Recognition}.

\bibitem[{Deng et~al.(2009)Deng, Dong, Sochera, Li, Li, and Fei-Fei}]{ImageNet}
Deng, J.; Dong, W.; Sochera, R.; Li, L.-J.; Li, K.; and Fei-Fei, L. 2009.
\newblock Imagenet: A large-scale hierarchical image database.
\newblock In \emph{Proceedings of the IEEE/CVF Conference on Computer Vision
  and Pattern Recognition}.

\bibitem[{Doan et~al.(2021)Doan, Lao, Zhao, and Li}]{LIRA}
Doan, K.; Lao, Y.; Zhao, W.; and Li, P. 2021.
\newblock LIRA: Learnable, Imperceptible and Robust Backdoor Attacks.
\newblock In \emph{Proceedings of the IEEE/CVF International Conference on
  Computer Vision}.

\bibitem[{Dosovitskiy et~al.(2021)Dosovitskiy, Beyer, Kolesnikov, Weissenborn,
  Zhai, Unterthiner, Dehghani, Minderer, Heigold, Gelly, Uszkoreit, and
  Houlsby}]{vit}
Dosovitskiy, A.; Beyer, L.; Kolesnikov, A.; Weissenborn, D.; Zhai, X.;
  Unterthiner, T.; Dehghani, M.; Minderer, M.; Heigold, G.; Gelly, S.;
  Uszkoreit, J.; and Houlsby, N. 2021.
\newblock An Image is Worth 16x16 Words: Transformers for Image Recognition at
  Scale.
\newblock In \emph{International Conference on Learning Representations}.

\bibitem[{Gao et~al.(2023)Gao, Bai, Gu, Yang, and Xia}]{ASD}
Gao, K.; Bai, Y.; Gu, J.; Yang, Y.; and Xia, S.-T. 2023.
\newblock Backdoor Defense via Adaptively Splitting Poisoned Dataset.
\newblock In \emph{Proceedings of the IEEE/CVF Conference on Computer Vision
  and Pattern Recognition}.

\bibitem[{Grill et~al.(2020)Grill, Strub, Altch\'{e}, Tallec, Richemond,
  Buchatskaya, Doersch, Pires, Guo, Azar, Piot, Kavukcuoglu, Munos, and
  Valko}]{BYOL}
Grill, J.-B.; Strub, F.; Altch\'{e}, F.; Tallec, C.; Richemond, P.~H.;
  Buchatskaya, E.; Doersch, C.; Pires, B.~A.; Guo, Z.~D.; Azar, M.~G.; Piot,
  B.; Kavukcuoglu, K.; Munos, R.; and Valko, M. 2020.
\newblock Bootstrap Your Own Latent a New Approach to Self-Supervised Learning.
\newblock In \emph{Advances in Neural Information Processing Systems}.

\bibitem[{Gu, Dolan-Gavitt, and Garg(2017)}]{BadNets}
Gu, T.; Dolan-Gavitt, B.; and Garg, S. 2017.
\newblock BadNets: Identifying Vulnerabilities in the Machine Learning Model
  Supply Chain.
\newblock \emph{arXiv preprint arXiv:1708.06733}.

\bibitem[{Guo et~al.(2023)Guo, Li, Chen, Guo, Sun, and Liu}]{SCALEUP}
Guo, J.; Li, Y.; Chen, X.; Guo, H.; Sun, L.; and Liu, C. 2023.
\newblock SCALE-UP: An Efficient Black-box Input-level Backdoor Detection via
  Analyzing Scaled Prediction Consistency.
\newblock In \emph{International Conference on Learning Representations}.

\bibitem[{Hayase and Oh(2023)}]{hayase2023fewshot}
Hayase, J.; and Oh, S. 2023.
\newblock Few-shot Backdoor Attacks via Neural Tangent Kernels.
\newblock In \emph{International Conference on Learning Representations}.

\bibitem[{He et~al.(2016)He, Zhang, Ren, and Sun}]{he2016deep}
He, K.; Zhang, X.; Ren, S.; and Sun, J. 2016.
\newblock Deep residual learning for image recognition.
\newblock In \emph{Proceedings of the IEEE Conference on Computer Vision and
  Pattern Recognition}.

\bibitem[{Ho, Jain, and Abbeel(2020)}]{DDPM}
Ho, J.; Jain, A.; and Abbeel, P. 2020.
\newblock Denoising Diffusion Probabilistic Models.
\newblock In \emph{Advances in Neural Information Processing Systems}.

\bibitem[{Huang et~al.(2022)Huang, Li, Wu, Qin, and Ren}]{DBD}
Huang, K.; Li, Y.; Wu, B.; Qin, Z.; and Ren, K. 2022.
\newblock Backdoor Defense via Decoupling the Training Process.
\newblock In \emph{International Conference on Learning Representations}.

\bibitem[{Jebreel, Domingo-Ferrer, and Li(2023)}]{layer_wise}
Jebreel, N.~M.; Domingo-Ferrer, J.; and Li, Y. 2023.
\newblock Defending Against Backdoor Attacks by Layer-wise Feature Analysis.
\newblock In \emph{Advances in Knowledge Discovery and Data Mining}.

\bibitem[{Jiang et~al.(2023)Jiang, Li, Xu, and Zhang}]{Jiang_2023_CVPR}
Jiang, W.; Li, H.; Xu, G.; and Zhang, T. 2023.
\newblock Color Backdoor: A Robust Poisoning Attack in Color Space.
\newblock In \emph{Proceedings of the IEEE/CVF Conference on Computer Vision
  and Pattern Recognition}.

\bibitem[{Krizhevsky, Hinton et~al.(2009)}]{cifar}
Krizhevsky, A.; Hinton, G.; et~al. 2009.
\newblock Learning multiple layers of features from tiny images.

\bibitem[{Li et~al.(2023)Li, Liu, Chen, Xie, Zhang, and
  Liu}]{li-etal-2023-multi-target}
Li, Y.; Liu, S.; Chen, K.; Xie, X.; Zhang, T.; and Liu, Y. 2023.
\newblock Multi-target Backdoor Attacks for Code Pre-trained Models.
\newblock In \emph{Proceedings of the 61st Annual Meeting of the Association
  for Computational Linguistics (Volume 1: Long Papers)}.

\bibitem[{Li et~al.(2021)Li, Lyu, Koren, Lyu, Li, and Ma}]{ABL}
Li, Y.; Lyu, X.; Koren, N.; Lyu, L.; Li, B.; and Ma, X. 2021.
\newblock Anti-Backdoor Learning: Training Clean Models on Poisoned Data.
\newblock In \emph{Advances in Neural Information Processing Systems}.

\bibitem[{Liu et~al.(2023)Liu, Li, Wang, Hu, Ye, Jin, Wu, and Xiao}]{teco}
Liu, X.; Li, M.; Wang, H.; Hu, S.; Ye, D.; Jin, H.; Wu, L.; and Xiao, C. 2023.
\newblock Detecting Backdoors During the Inference Stage Based on Corruption
  Robustness Consistency.
\newblock In \emph{Proceedings of the IEEE/CVF Conference on Computer Vision
  and Pattern Recognition}.

\bibitem[{Liu, Bailey, and Lu(2020)}]{Refool}
Liu, X., Yunfeand~Ma; Bailey, J.; and Lu, F. 2020.
\newblock Reflection Backdoor: A Natural Backdoor Attack on Deep Neural
  Networks.
\newblock In \emph{Proceedings of the European Conference on Computer Vision}.

\bibitem[{Liu et~al.(2017)Liu, Ma, Aafer, Lee, Zhai, Wang, and Zhang}]{Trojan}
Liu, Y.; Ma, S.; Aafer, Y.; Lee, W.-C.; Zhai, J.; Wang, W.; and Zhang, X. 2017.
\newblock Trojaning attack on neural networks.
\newblock \emph{Annual Network And Distributed System Security Symposium}.

\bibitem[{Ma et~al.(2022)Ma, Li, Gao, Abuadbba, Zhang, Fu, Kim, Al-Sarawi,
  Surya, and Abbott}]{ma2022dangerous}
Ma, H.; Li, Y.; Gao, Y.; Abuadbba, A.; Zhang, Z.; Fu, A.; Kim, H.; Al-Sarawi,
  S.~F.; Surya, N.; and Abbott, D. 2022.
\newblock Dangerous cloaking: Natural trigger based backdoor attacks on object
  detectors in the physical world.
\newblock \emph{arXiv preprint arXiv:2201.08619}.

\bibitem[{Nguyen and Tran(2020)}]{Dynamic}
Nguyen, T.~A.; and Tran, A. 2020.
\newblock Input-Aware Dynamic Backdoor Attack.
\newblock In \emph{Advances in Neural Information Processing Systems}.

\bibitem[{Nguyen and Tran(2021)}]{WaNet}
Nguyen, T.~A.; and Tran, A.~T. 2021.
\newblock WaNet - Imperceptible Warping-based Backdoor Attack.
\newblock In \emph{International Conference on Learning Representations}.

\bibitem[{Paszke et~al.(2019)Paszke, Gross, Massa, Lerer, Bradbury, Chanan,
  Killeen, Lin, Gimelshein, Antiga, Desmaison, Kopf, Yang, DeVito, Raison,
  Tejani, Chilamkurthy, Steiner, Fang, Bai, and
  Chintala}]{NEURIPS2019_bdbca288}
Paszke, A.; Gross, S.; Massa, F.; Lerer, A.; Bradbury, J.; Chanan, G.; Killeen,
  T.; Lin, Z.; Gimelshein, N.; Antiga, L.; Desmaison, A.; Kopf, A.; Yang, E.;
  DeVito, Z.; Raison, M.; Tejani, A.; Chilamkurthy, S.; Steiner, B.; Fang, L.;
  Bai, J.; and Chintala, S. 2019.
\newblock PyTorch: An Imperative Style, High-Performance Deep Learning Library.
\newblock In \emph{Advances in Neural Information Processing Systems}.

\bibitem[{Qi et~al.(2023)Qi, Xie, Li, Mahloujifar, and
  Mittal}]{qi2023revisiting}
Qi, X.; Xie, T.; Li, Y.; Mahloujifar, S.; and Mittal, P. 2023.
\newblock Revisiting the Assumption of Latent Separability for Backdoor
  Defenses.
\newblock In \emph{International Conference on Learning Representations}.

\bibitem[{Redmon and Farhadi(2018)}]{redmon2018yolov3}
Redmon, J.; and Farhadi, A. 2018.
\newblock Yolov3: An incremental improvement.
\newblock \emph{arXiv preprint arXiv:1804.02767}.

\bibitem[{Saha et~al.(2022)Saha, Tejankar, Koohpayegani, and
  Pirsiavash}]{Saha_2022_CVPR}
Saha, A.; Tejankar, A.; Koohpayegani, S.~A.; and Pirsiavash, H. 2022.
\newblock Backdoor Attacks on Self-Supervised Learning.
\newblock In \emph{Proceedings of the IEEE/CVF Conference on Computer Vision
  and Pattern Recognition}.

\bibitem[{Shi et~al.(2023)Shi, Liu, Zhou, and Sun}]{BadGPT}
Shi, J.; Liu, Y.; Zhou, P.; and Sun, L. 2023.
\newblock BadGPT: Exploring Security Vulnerabilities of ChatGPT via Backdoor
  Attacks to InstructGPT.
\newblock \emph{Network and Distributed System Security Symposium}.

\bibitem[{Tan and Shokri(2020)}]{9230390}
Tan, T. J.~L.; and Shokri, R. 2020.
\newblock Bypassing Backdoor Detection Algorithms in Deep Learning.
\newblock In \emph{IEEE European Symposium on Security and Privacy}.

\bibitem[{Turner, Tsipras, and Madry(2019)}]{CL}
Turner, A.; Tsipras, D.; and Madry, A. 2019.
\newblock Clean-label backdoor attacks.
\newblock \emph{https://people.csail.mit.edu/madry/lab/}.

\bibitem[{Wang et~al.(2022)Wang, Ding, Zhai, and Ma}]{NONE}
Wang, Z.; Ding, H.; Zhai, J.; and Ma, S. 2022.
\newblock Training with More Confidence: Mitigating Injected and Natural
  Backdoors During Training.
\newblock In \emph{Advances in Neural Information Processing Systems}.

\bibitem[{Wang, Zhai, and Ma(2022)}]{BppAttack}
Wang, Z.; Zhai, J.; and Ma, S. 2022.
\newblock BppAttack: Stealthy and Efficient Trojan Attacks Against Deep Neural
  Networks via Image Quantization and Contrastive Adversarial Learning.
\newblock In \emph{Proceedings of the IEEE/CVF Conference on Computer Vision
  and Pattern Recognition}.

\bibitem[{Weng, Lee, and Wu(2020)}]{NEURIPS2020_8b406655}
Weng, C.-H.; Lee, Y.-T.; and Wu, S.-H.~B. 2020.
\newblock On the Trade-off between Adversarial and Backdoor Robustness.
\newblock In \emph{Advances in Neural Information Processing Systems}.

\bibitem[{Yuan et~al.(2023)Yuan, Zhou, Zou, and Cheng}]{Yuan_2023_CVPR}
Yuan, Z.; Zhou, P.; Zou, K.; and Cheng, Y. 2023.
\newblock You Are Catching My Attention: Are Vision Transformers Bad Learners
  Under Backdoor Attacks?
\newblock In \emph{Proceedings of the IEEE/CVF Conference on Computer Vision
  and Pattern Recognition}.

\bibitem[{Zhang(2004)}]{SGD}
Zhang, T. 2004.
\newblock Solving Large Scale Linear Prediction Problems Using Stochastic
  Gradient Descent Algorithms.
\newblock In \emph{Proceedings of the International Conference on Machine
  Learning}.

\end{thebibliography}

\clearpage

\renewcommand\thesection{\Alph{section}}
\setcounter{section}{0}

\section{Algorithm outline}
\label{ap:algorithm}
With these newly proposed stages, we can summarize our PIPD method in Algorithm \ref{alg:algorithm}. The algorithm, in lines 1-4, illustrates the stage of obtaining poisoned subset and benign subset. In lines 6-8, we use these two subsets to execute a single-round isolation. Subsequently, the predicted benign set $\hat{\mathcal{D}}_b$ is unioned with the benign subset $\mathcal{P}_b$ for the subsequent iteration of isolation. Upon completion of the isolation stage, as presented in lines 11-13, we train the model on the predicted benign set and carry out an selective training process on the identified poisoned set.

\begin{algorithm}[!htbp]
\caption{Proposed PIPD}
\label{alg:algorithm}
\textbf{Input}: Network $f$ with parameter $\theta$; poisoned dataset $\mathcal{D}$; pre-isolation epochs $N_{1}$ and percentage \emph{p, q}; progressive iteration number $T$; selective training epochs $N_{2}$.\\
\textbf{Output}: Predicted $\hat{\mathcal{D}}_p^T$ and $\hat{\mathcal{D}}_b^T$; well-trained $f$ with $\theta^{*}$.    
\begin{algorithmic}[1]
\Statex \emph{\#Pre-isolation Phase}:
\For{t = 1 to $N_{1}$}
\State Update $\theta$ using LGA loss. \Comment{Eq. \eqref{eq:ABL_loss_eq}}
\EndFor
\State Initialize $\mathcal{P}_b^{1}$ from $\mathcal{D}$ with the highest \emph{q}\% loss.
\State Initialize $\mathcal{P}_p^{1}$ from $\mathcal{D}$ with the lowest \emph{p}\% loss.
\Statex \emph{\#Progressive Isolation Phase}:
\For{t = 1 to $T-1$}
\State Identifying $d_i^{l}$. \Comment{Eq. \eqref{eq:discrepancy_measure}}
\State Aggregating $\mathcal{S}$. \Comment{Eq. \eqref{eq:scores}}
\State Clustering $\mathcal{D}$ to $\hat{\mathcal{D}}_b^{t}$ and $\hat{\mathcal{D}}_p^{t}$ based on $\mathcal{S}$. \Comment{Eq. \eqref{eq:cluster}}
\State Update $\mathcal{P}_b^{t+1}$. \Comment{Eq. \eqref{eq:enlarge}}
\EndFor
\Statex \emph{\#Selective Training Phase}:
\For{t = 1 to $N_{2}$}
\State Update $\theta$ using $\hat{\mathcal{D}}_b^T$ and $\hat{\mathcal{D}}_p^T$. \Comment{Eq. \eqref{eq:unlearn_schema_1}}
\EndFor
\end{algorithmic}
\end{algorithm}

\section{Implementation details}
\label{sec:Implementation details}
In summary, we use the framework PyTorch \cite{NEURIPS2019_bdbca288} to implement all the experiments. Note that the experiments on CIFAR-10 and ImageNet are run on a NVIDIA Tesla V100 GPU with 32GB memory.

\subsection{Datasets and DNN models}\label{ap:dataset_details}
The details of datasets and DNN models in our experiments are summarized in Table \ref{datasets and models}. Specially, we randomly choose 30 classes from ImageNet to construct a subset due to the limitation of the computational time and costs.

\begin{table}[!htbp]
\begin{center}
\caption{Summary of datasets and DNN models in our experiments.}
\label{datasets and models}
\footnotesize
\setlength{\tabcolsep}{0.3mm}{
\scalebox{0.9}{
\begin{tabular}{cccccc}
\toprule[0.68pt]
\multicolumn{1}{c}{\multirow{2}{*}{Dataset}} & \multicolumn{1}{c}{\multirow{2}{*}{\# Input size}}  & \multicolumn{1}{c}{\multirow{2}{*}{\# Classes}} & \multicolumn{1}{c}{\# Training}  & \multicolumn{1}{c}{\# Testing} & \multicolumn{1}{c}{\multirow{2}{*}{Models}}\\
 &  &  &  images &  images & \\
\midrule
CIFAR-10 & 3 $\times$ 32 $\times$ 32 & 10 & 50000 & 10000 & ResNet-18  \\
\midrule
ImageNet & 3 $\times$ 224 $\times$ 224 & 30 & 38859 & 1500 & ResNet-18 \\
\bottomrule[0.68pt]
\end{tabular}}}
\end{center}
\end{table}

\subsection{Attack setups}
\noindent \textbf{Poisoned dataset setups.} In pursuit of achieving fairness for evaluating isolation-based defense methods, we use the open-source PyTorch code from\footnote{https://github.com/vtu81/backdoor-toolbox} to create the poisoned CIFAR-10, storing the poisoned dataset rather than poisoning it during the training process. We discern that some implementations involve first retrieve a batch of benign samples then add triggers, where every sample has been or had been previously poisoned. This could heavily affect the defense result, especially for isolation-based method in terms of evaluating isolation quality and defense performance. As for the ImageNet, we adopt the code from\footnote{https://github.com/VITA-Group/Trap-and-Replace-Backdoor-Defense} to create and store poisoned ImageNet before training.

\noindent \textbf{Training setups.} On the CIFAR-10 \cite{cifar} dataset, we perform backdoor attacks on ResNet-18 \cite{he2016deep} for 200 epochs with batch size 128. We adopt the stochastic gradient descent (SGD) \cite{SGD} optimizer with a learning rate 0.1, momentum 0.9, weight decay $5\times10^{-4}$. The learning rate is divided by 10 at epoch 100 and 150. On the ImageNet \cite{ImageNet} dataset, we train ResNet-18 for 90 epochs with batch size 256, where the SGD optimizer is initialized with a learning rate 0.1, momentum 0.9, and weight decay $10^{-4}$. The learning rate is decreased by a factor of 10 at epoch 30 and 60. The image resolution is resized to $224\times224\times3$ before attaching the trigger. \par

\subsection{Attack Details} \label{ap:attack_details}
We mainly consider 9 state-of-the-art backdoor attacks, including: BadNets \cite{BadNets}, Trojan \cite{Trojan}, Blended \cite{Blend}, Dynamic \cite{Dynamic}, WaNet \cite{WaNet}, SIG \cite{SIG}, Refool \cite{Refool}, and CL \cite{CL}. In addition, we evaluate one recently proposed adaptive attack termed Adaptive-Blend (A-Patch) \cite{qi2023revisiting}.

We use two standard data augmentation techniques (horizontal flip and random crop with padding $4 \times 4$ for CIFAR-10 and $16 \times 16$ for ImageNet) during model training. We follow the default settings suggested in the original papers and the open-source codes for most attacks; these include the trigger pattern, trigger size, and backdoor label. The backdoor label of all attacks is set to class 0. Table \ref{tab:attacks_overview} summarizes the detailed settings of these attacks.

\begin{table}[!htbp]
\centering
\caption{Attack settings of 9 backdoor attacks. ASR (\%): attack success rate; CA (\%): clean accuracy.}
\label{tab:attacks_overview}
\scalebox{0.8}{
\begin{tabular}{cccc}
\toprule
Attacks & Trigger Pattern & Target Label & Poisoning Rate \\ 
\midrule
BadNets & Pixel Patch & 0 & 0.05 \\
Trojan & Reversed Watermark & 0 & 0.05  \\
Blended & Random Pixel & 0 & 0.05  \\
Refool & Reflection-injected & 0 & 0.05  \\
Dynamic & Mask Generator & 0 & 0.05 \\
SIG & Sinusoidal Signal & 0 & 0.08 \\
CL & Label-consistent & 0 & 0.08 \\
WaNet & Optimization-based & 0 & 0.05 \\
A-Blend & Mixer Construction & 0 & 0.05 \\
\bottomrule
\end{tabular}}
\end{table}

\subsection{Competing Methods Details} \label{ap:defense_details}
We experiment with four training-time defense methods: ABL \cite{ABL}, DBD \cite{DBD}, NONE \cite{NONE}, and ASD \cite{ASD}, and two backdoor detection methods: DBAFLA \cite{layer_wise} and SCALE-UP \cite{SCALEUP}. When evaluating isolation result on DBALFA, we use 10\% benign test set, as specified in their work. Regarding DBD and ASD, when evaluating the isolation quality, we will mark the data determined as unlabelled by them as poisoned, since the labels of this part of data were lost during the training process. 

We use the open-source PyTorch codes for ABL\footnote{https://github.com/bboylyg/ABL}, DBD\footnote{https://github.com/SCLBD/DBD}, ASD\footnote{https://github.com/KuofengGao/ASD}, NONE\footnote{https://github.com/RU-System-Software-and-Security/NONE}, DBAFLA\footnote{https://github.com/najeebjebreel/dbalfa}, and SCALE-UP\footnote{https://github.com/junfenggo/scale-up} to reproduce the results of them. 

\subsection{Metric Details}\label{ap:metric_details}
We use four metrics for evaluation: 1) Clean Accuracy (CA) - the model's performance on the clean test set; 2) Attack Success Rate (ASR) - the model's performance on the poisoned test set; 3) True Positive Rate (TPR) - calculated as the ratio of isolated poisoned inputs to the total number of poisoned inputs, and 4) False Positive Rate (FPR) - the ratio of benign inputs incorrectly isolated as poisoned to the total number of benign inputs. These metrics were adopted from the DBALFA. Higher TPR and CA values, and lower FPR and ASR values generally suggest a more efficient isolation process and an improved training-time defense method, respectively.

\section{Additional Experimental Results of PIPD}
\subsection{The Impact of Poison Rates}
\label{ap:poison-rate}
This section investigates the effectiveness of our method against varying poisoning rates. In the following experiments, we set the poisoning rates for five distinct attacks at 5\%, 10\%, 15\%, 20\%, and 30\%. We use a ResNet-18 and CIFAR-10 dataset for the illustration, the results are detailed in Table.~\ref{tab:result-of-poison-rate}. As the poison rate gradually increases, the TPR begins to decline while the FPR starts to rise. This suggests that the difficulty of isolation intensifies with the escalation of the poison rate.

\begin{table}[!tbp]
	\centering
        \setlength\tabcolsep{3pt}
	\scalebox{0.9}{
	\begin{tabular}{c|rrrr}
		\toprule
            \addlinespace[0pt]
		\multicolumn{1}{c|}{\tabincell{c}{$\eta$}} & BadNets & Blended & Dynamic & Trojan \\
            \hline
		  5\% & 100.00/0.00 & 100.00/0.00 & 100.00/0.00 & 100.00/0.00 \\
            10\% & 100.00/0.00 & 99.62/0.23 & 99.94/0.00 & 100.00/0.00\\
            15\% & 100.00/0.00 & 99.30/0.23 & 99.89/0.00 & 100.00/0.07\\
            20\% & 100.00/0.53 & 99.04/0.61 & 99.88/0.00 & 100.00/0.83\\
            30\% & 99.63/1.00 & 98.66/0.77 & 99.88/0.00 & 98.48/1.27\\
		\addlinespace[-0.22em]
            \bottomrule[0.68pt]
	\end{tabular}}
	\caption{TPR/FPR isolation results (\%) of our PIPD on attacks with different poison rates.}
	\label{tab:result-of-poison-rate}
\end{table}

\subsection{The Impact of Clustering}
\label{ap:cluster}

This part of our study evaluates the impact of clustering algorithms used in the isolation process on the quality of isolation. Again, we experiment on the ResNet-18 trained on poisoned CIFAR-10 with 5\% poison rate attacked by various attacks. We adopt different clustering algorithms replacing the Fisher-Jenks algorithm. We list the TPR and FPR by using K-means, Bisecting K-Means and Mini-Batch K-Means, in addition to our default Fisher-Jenks in Table.~\ref{tab:different_clustering}. As can be noticed, all TPR and FPR results with different clustering algorithms are similar. This implies that the isolation quality is not sensitive to the clustering algorithms adopted.

\begin{table}[t]
\footnotesize
\setlength{\tabcolsep}{0.9mm}{
\begin{tabular}{l|rrrrrrrr}
\toprule[0.68pt]
\addlinespace[0pt]
\multicolumn{1}{c|}{\multirow{2}{*}{Method}} & \multicolumn{2}{c}{BadNets} & \multicolumn{2}{c}{Blended} & \multicolumn{2}{c}{Trojan} & \multicolumn{2}{c}{Dynamic} \\ \cline{2-9} 
\multicolumn{1}{c|}{} & TPR & FPR & TPR & FPR & TPR & FPR & TPR & FPR\\ \hline
Fisher-Jenks & 100.0 & 0.00	& 100.0 & 0.00 & 100.0 & 0.00 & 100.0 & 0.00\\
K-means & 100.0 & 0.00 & 99.88 & 0.65 & 100.0 & 0.05 & 100.0 & 0.41\\
B-KMeans & 100.0 & 0.00 & 99.76 & 0.28	& 100.0 & 0.77 & 100.0 & 0.00\\
MB-KMeans & 100.0 & 0.00 & 99.80 & 0.10 & 100.0 & 0.01 & 99.96 & 0.00  \\ 
\addlinespace[-0.22em]
\bottomrule[0.68pt]
\end{tabular}}
\caption{The TPR(\%) and FPR(\%) on CIFAR-10 of our PIPD using different clustering methods.}
\label{tab:different_clustering}
\end{table}

\end{document}